# A Wide-Input 0.25-μm BCD LDO with Dual-Stage Amplifier and Active Ripple Cancellation for High PSRR and Fast Transient Response

Yi Zhang, *Student Member, IEEE*, Zhuolong Chen, *Student Member, IEEE*, Zhenghao Xu, *Student Member, IEEE*, and Yujin He, *Student Member, IEEE*

*Abstract*—Demand for on-chip low-dropout regulators (LDOs) with both high power-supply rejection ratio (PSRR) and fast transient response is growing as system-on-chip (SoC) integration increases. However, conventional LDO architectures face difficulty achieving these performance metrics simultaneously over wide input voltage ranges. This paper presents a wide-input linear regulator implemented in 0.25-μm BCD technology that attains high PSRR and swift load-transient performance while maintaining low quiescent current. The proposed LDO employs a dual-stage error amplifier architecture and active ripple cancellation along both the power path and the error amplifier's supply to significantly enhance PSRR across frequency. An adaptive fast feedback branch together with an on-chip frequency compensation network is introduced to accelerate transient response without compromising stability. A two-stage PSRR analytical model and a three-frequency-band PSRR interpretation framework are developed to guide the design. Cadence Spectre simulations of the 14 V-output LDO (with input up to 48 V) demonstrate a −75 dB low-frequency PSRR, and during a 50 μA–4 mA load step the output voltage droop is kept under 0.65 V with recovery within 16 μs. These results validate the effectiveness of the proposed architecture and analysis, indicating that the design meets the stringent requirements of analog/RF SoCs and portable electronics.

## I. Introduction

WITH the ongoing integration of power management into system-on-chip (SoC) designs, low-dropout regulators (LDOs) are increasingly embedded on the same die as their load circuits. On-chip integration not only reduces cost and size but also minimizes parasitic noise coupling from package interconnects. High power-supply rejection (PSR) has become essential for LDOs that feed noise-sensitive analog and RF blocks, since a well-regulated, clean supply is required to shield these circuits from switching noise and ripple on the supply rails 0. At the same time, modern SoCs and portable devices impose rapidly varying load currents as digital blocks transition between modes, causing sudden output voltage dips or surges. Regulators must respond swiftly to such load transients, because most digital and mixed-signal circuits do not tolerate large supply glitches [2]. As a result, the research focus for LDO design has shifted from simply achieving low dropout and high load current toward optimizing PSRR and transient response performance. Achieving both high PSRR and fast transient settling concurrently is now a critical requirement for advanced SoC power management.

Conventional LDO architectures struggle to simultaneously meet the above demands due to fundamental trade-offs. High DC gain and bandwidth in the error amplifier are necessary to improve PSR, yet even then traditional LDOs often exhibit poor PSRR at higher frequencies (hundreds of kHz and above) [3]. Various techniques have been explored to boost PSRR. For example, adding an RC filter at the LDO output or cascading two regulation stages can improve PSR at the expense of increased dropout voltage. Using an NMOS pass transistor with a charge-pump boosted gate bias is another approach to extend bandwidth and raise PSRR, but this tends to increase quiescent power draw and design complexity [4]. Feed-forward ripple cancellation techniques, which inject a replica of the input ripple into the pass device's control node, have proven effective in significantly enhancing PSRR over a broad frequency range 0. El-Nozahi *et al.* demonstrated that such feed-forward paths can substantially cancel supply noise at the LDO output [4]. However, this method requires a wide-bandwidth auxiliary amplifier to track high-frequency supply ripple, incurring additional bias current and circuit overhead. Another strategy is to employ multiple feedback loops or multi-stage amplifiers to increase open-loop gain and widen the control bandwidth, which can improve both PSRR and transient speed.

This paragraph of the first footnote will contain the date on which you submitted your paper for review, which is populated by IEEE. It is IEEE style to display support information, including sponsor and financial support acknowledgment, here and not in an acknowledgment section at the end of the article. For example, "This work was supported in part by the U.S. Department of Commerce under Grant 123456." The name of the corresponding author appears after the financial information, e.g. *(Corresponding author: Second B. Author).* Here you may also indicate if authors contributed equally or if there are co-first authors.

Yi Zhang is with School of Advanced Manufacturing, Fuzhou University, Quanzhou 362251, China (e-mail: 852203429@fzu.edu.cn ).

Zhuolong Chen is with School of Advanced Manufacturing, Fuzhou University, Quanzhou 362251, China (e-mail: 203900129@qq.com ).

Zhenghao Xu is with School of Advanced Manufacturing, Fuzhou University, Quanzhou 362251, China (e-mail: 1806758022@qq.com ).

Yujin He is with School of Advanced Manufacturing, Fuzhou University, Quanzhou 362251, China (e-mail: 1021771427@qq.com ).

Digital Object Identifier:

Unfortunately, introducing extra gain stages and loops raises power consumption and complicates frequency compensation, risking stability issues. In high-voltage LDO designs (such as for automotive applications), cascoding pass transistors and using charge-pump biasing have been used to achieve wide input operation and better PSR by increasing output impedance. Yet, these solutions often suffer from reduced headroom (higher dropout) and sensitivity to process or supply variations. Many wide-input LDOs also resort to large off-chip capacitors or pre-regulator circuits for noise suppression, which can drastically lower the LDO's bandwidth and slow its transient response 0. In summary, prior approaches typically address either PSR or transient performance, but achieving excellent performance in both domains simultaneously—while also maintaining low quiescent current and wide input range capability—remains challenging.

To meet these challenges, this paper proposes a new LDO architecture that pursues high PSRR and fast transient response concurrently. The design targets a wide input voltage range and low quiescent power, suitable for on-chip regulation in mixed-signal SoCs. The main contributions of this work are summarized as follows:

1. **Dual-Stage Error Amplifier Architecture:** A two-stage error amplifier is employed to provide high DC gain across a wide output voltage range, improving low-frequency PSRR and ensuring sufficient loop gain even at high input voltages. The dual-stage topology is optimized for stability via internal compensation, enabling the regulator to handle a wide input (tested up to 48 V) while maintaining regulation accuracy.
2. **Active Ripple Cancellation for PSRR Enhancement:** The LDO actively cancels supply ripple by feeding ripple components into both the power transistor gate and the error amplifier's supply node. This active ripple cancellation along the power path and amplifier supply significantly boosts the PSRR over a broad frequency spectrum, attenuating both low-frequency supply noise and high-frequency ripple coupling. The approach mitigates the dominant noise coupling mechanisms without requiring large external components.
3. **Adaptive Fast Feedback Branch for Transient Improvement:** An auxiliary fast feedback loop is introduced in parallel with the main error amplifier to accelerate the response to load current changes. This adaptive fast branch senses rapid output deviations and provides a direct drive to the pass device, drastically reducing overshoot and undershoot during transients. Combined with an on-chip frequency compensation network, this technique improves the slew rate of the control loop and shortens the settling time after load steps.
4. **Prototype Implementation and Performance:** The proposed LDO is implemented in a 0.25-μm BCD process as a 14 V output regulator that accommodates input voltages from 5 V up to 48 V. Cadence Spectre simulations verify that the LDO achieves a **−75 dB** PSRR at DC (low frequency) and maintains above −30 dB PSRR at hundreds of kHz frequencies. For a load current step of 50 μA to 4 mA (a full-scale heavy load transition), the output voltage deviation is limited to **0.65 V** (droop) and recovers to within 2% of nominal in under **16 μs**. The LDO draws a low quiescent current under light-load conditions, thanks to the power-efficient biasing of the two-stage amplifier and adaptive loop.
5. **PSRR Analysis Model and Design Framework:** In addition to the circuit implementation, the paper contributes a theoretical two-stage PSRR analysis model that quantifies the PSR contributions of the bandgap reference, error amplifier, and pass device. Using this model, a three-frequency-band interpretation of the LDO's PSRR behavior is formulated, dividing the PSRR response into low-, mid-, and high-frequency regions with distinct limiting factors. This framework provides insight into the PSRR improvement mechanisms (such as ripple cancellation) and is validated against the simulation results, which show close agreement with the predicted three-band PSRR characteristics.

In summary, the proposed 0.25-μm BCD LDO design achieves a combination of wide input range, high PSRR, fast transient response, and low quiescent current that is difficult to obtain using conventional techniques. The presented architecture and analytical approach can help enable integrated regulators for analog and RF sub-systems in SoCs and mobile devices that demand clean and stable supply voltages under dynamic operating conditions. The following sections detail the LDO circuit implementation and theoretical analysis, present simulation results that verify the performance improvements, and compare the design with state-of-the-art LDO solutions.

## II. Overview of Low Dropout Linear Regulators

In consumer electronics, if the circuitry used requires a power supply with high noise and ripple rejection, transient calibration, and output state self-test, then LDO is a very appropriate choice.

Fig. 1 shows a typical LDO architecture, where the error amplifier and the power output stage form the main negative feedback loop of the LDO, the reference circuit provides the reference clamping voltage, and the commonly used auxiliary circuits such as disconnect circuits, short-circuit/over-temperature protection circuits, frequency compensation circuits, and so on are used to improve the specific performance of the LDO. In this report, in order to make the design of the LDO circuits more intuitive, the auxiliary module only includes the necessary frequency compensation circuits. The auxiliary modules in this report only include the necessary frequency compensation circuits to make LDO circuit design more intuitive. In order to facilitate the analysis of the LDO operating principle, Fig. 1 shows its typical structure.

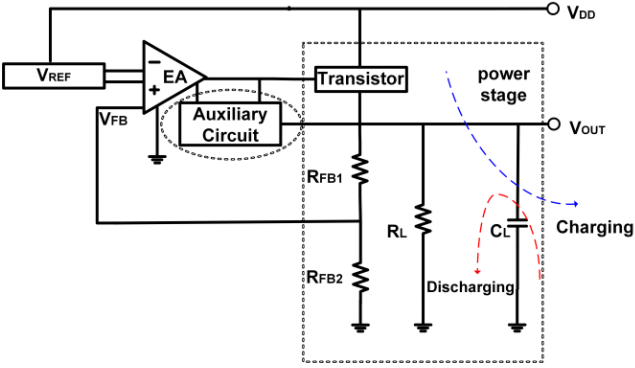

**Fig. 1.** Typical LDO Architecture Components

The specific working principle of the LDO is as follows: When the power supply $V_{DD}$ is turned on, the bandgap reference startup circuit rapidly activates its core circuitry, generating a stable reference voltage $V_{REF}$ that is unaffected by changes in voltage, temperature, or manufacturing process. During the power-up process, the feedback voltage $V_{FB}$ is unsteady. the voltage and $V_{REF}$ to do the difference and through the error amplifier output to regulate the drain current of the power transistor, thereby controlling the charging and discharging of the output capacitor $C_L$. At steady state, $V_{FB}$ is clamped by $V_{REF}$, and the LDO output provides a voltage proportional to $V_{REF}$. When power supply ripple is applied as the LDO input, load current changes, or both occur simultaneously, assuming the output voltage is pulled down (i.e., $C_L$ discharges), the pulled-down output voltage first passes through the feedback network $V_{REF1}$, $V_{REF2}$, and error amplifier to be fed back to the power transistor's gate. This lowers the gate potential, increasing the power transistor's drain current, causing $C_L$ to begin charging and raising the output voltage, ultimately achieving stable voltage regulation.

## III. KEY PERFORMANCE PARAMETERS

With the development and advancement of SoC and consumer electronics such as mobile phones and laptops, the focus of LDO circuit research has shifted from low power consumption and high load current to optimizing high power supply rejection ratio (PSRR) and transient response performance. This report will provide a detailed introduction to the research methods and technologies related to LDO PSRR and transient response performance.

### A. LDO PSRR Research Methods

In addition to stabilizing the output voltage, one of the most important functions of an LDO is to isolate noise from the external power supply and provide clean power to the circuits it drives. For this reason, the PSRR of an LDO has become a key consideration in circuit design. To gain a more systematic understanding of the factors that influence PSRR performance, this report analyzes the factors affecting PSRR performance from two perspectives: the contribution of each LDO module to PSRR performance and the PSRR characteristics at different frequency bands.

### B. Two-Stage Operational Amplifier PSRR Analysis Method

Considering the contribution of the error amplifier and bandgap reference power supply ripple feedthrough to the overall LDO power supply ripple suppression, the LDO can be equivalently divided into two stages as follows:

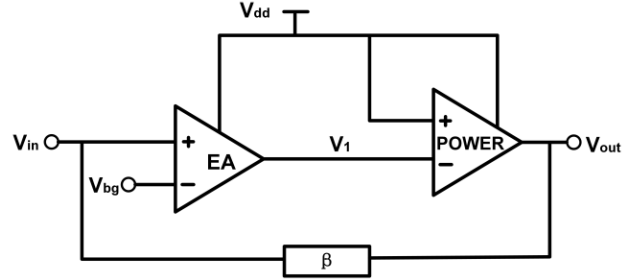

**Fig. 2.** LDO Two-Stage Op Amp Equivalent Diagrams

According to the equivalent circuit of the two-stage operational amplifier, the following equations can be listed:

$$\begin{cases} V_1 = PSRR_{EA}V_{dd} + (V_{in} - PSRR_{bg}V_{dd})A_1 \\ V_{out} = PSRR_{pow}V_{dd} + (V_{dd} - V_1)A_2 \\ V_{in} = \beta V_{out} \end{cases} \quad (1)$$

In the equation, $A_1$ and $A_2$ are the open-loop gains provided by the error amplifier and power stage, respectively. It should be noted that $A_2$ is positive at this point, where $PSRR_{EA} = V_{1\_Vdd}/V_{dd}$, $PSRR_{pow} = V_{pow\_Vdd}/V_{dd}$, $PSRR_{bg} = V_{bg\_Vdd}/V_{dd}$

Furthermore:

$$PSRR_{LDO} = \frac{V_{out}}{V_{dd}} = \frac{PSRR_{Bg}A_1A_2 - PSRR_{EA}A_2 + PSRR_{pow} + A_2}{1 + \beta A_1 A_2}$$

$$\approx \frac{PSRR_{bg}}{\beta} - \frac{PSRR_{EA}}{\beta A_1} + \frac{PSRR_{pow}}{\beta A_1 A_2} \quad (2)$$

(2) shows the contribution of the bandgap reference, error amplifier, and power stage to the LDO PSRR performance. The design can be optimized by canceling out the last two terms in this equation and increasing the ripple suppression capability of the bandgap reference to improve the overall PSRR performance of the LDO [5]. This equation provides a quantitative basis for optimizing the overall PSRR performance of the LDO.

### C. Three-band model PSRR analysis method

Traditional LDOs use external large output capacitors for frequency compensation and voltage regulation, with the output pole as the main pole. The LDO circuit in this report does not consider external capacitors. Therefore, based on this method and combined with the actual main pole position, the PSRR performance analysis is as follows:

As shown in Fig. 3, the PSRR transfer function can be regarded as the parallel combination of the impedance between the power supply and the regulator output and the impedance between the output and ground (open-loop output impedance to ground $Z_o = (Z_{Cout} + R_{ESR}) \parallel (R_1 + R_2)$ and the closed-loop equivalent output impedance $Z_{o\_reg} = (Z_o \parallel r_{ds})/(A_{o\_dc} \cdot \beta)$ introduced by negative feedback. The resistance voltage division relationship between these two impedances is represented by the parallel connection $Z_o \parallel Z_{o\_reg}$. From this model, the following equations can be derived:

$$PSRR = \frac{(Z_o \parallel Z_{o\_reg})}{r_{ds} + (Z_o \parallel Z_{o\_reg})} \quad (3)$$

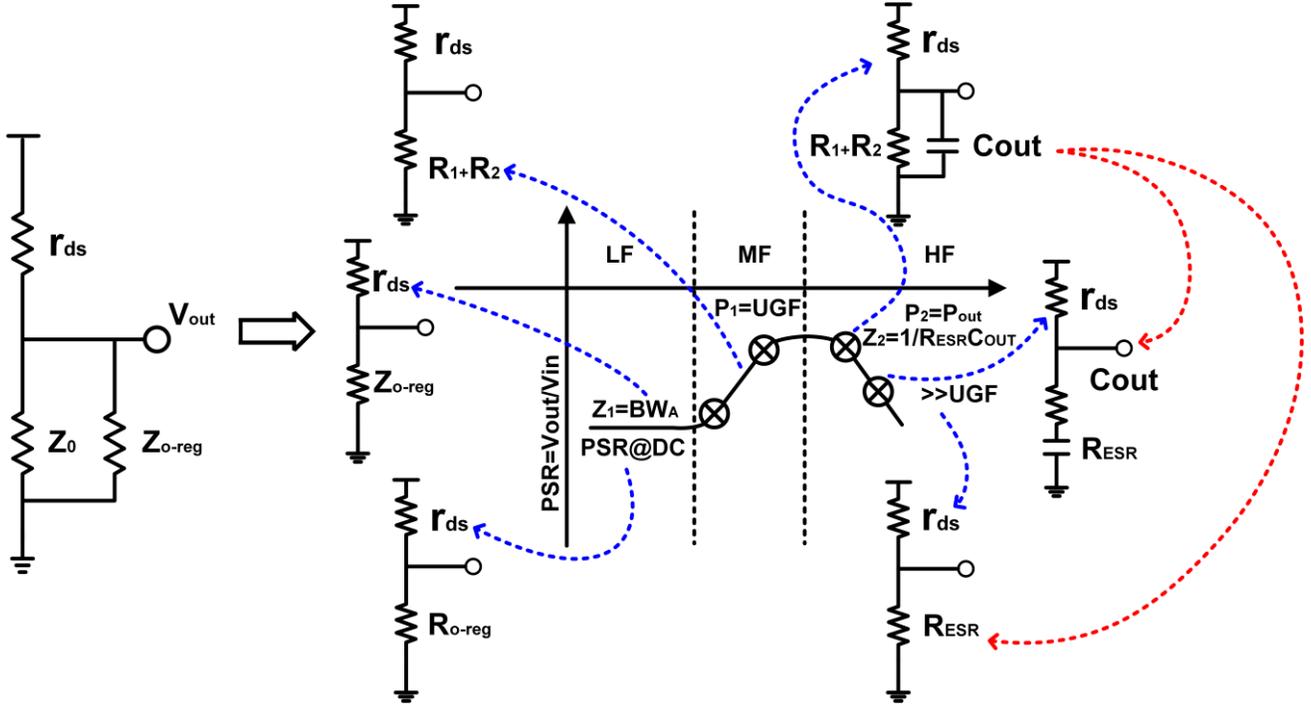

**Fig. 3.** Wide Frequency Range LDO PSRR Analytical Modeling

At low frequencies, the system loop gain is high, and the result of $Z_o$ and $Z_{o\_reg}$ being connected in parallel is $Z_{o\_reg}$, where the capacitive reactance of the load capacitor is infinite. Under normal circumstances, $r_{ds} \ll R_1 + R_2$, and (3) can be rewritten as:

$$PSRR_{DC} = \frac{\frac{r_{ds} \parallel (R_1+R_2)}{A_{o\_dc}\beta}}{r_{ds} + \frac{r_{ds} \parallel (R_1+R_2)}{A_{o\_dc}\beta}} \approx \frac{\frac{r_{ds}}{A_{o\_dc}\beta}}{r_{ds} + \frac{r_{ds}}{A_{o\_dc}\beta}} = \frac{1}{A_{o\_dc}\beta} \quad (4)$$

The results show that at low frequencies, LDO PSRR performance is proportional to the system loop gain.

When the power supply ripple frequency exceeds the system loop gain $-3\ dB$ bandwidth $BW_A$ (error amplifier output pole $P_{O-A}$), the loop gain begins to decrease, causing $Z_{o\_reg}$ to increase. However, since $Z_o$ and $Z_{o\_reg}$ are connected in parallel, $Z_{o\_reg}$ remains dominant. Meanwhile, the load capacitive reactance remains significant. The mid-frequency band is defined as the frequency band between the $BW_A$ and the loop gain unity gain bandwidth $GBW$.

The mid-frequency band PSRR can be rewritten using (3) as:

$$PSRR_{BW_A < f < GBW} \approx \frac{Z_{o\_reg}}{r_{ds} + Z_{o\_reg}}$$

$$= \frac{1 + \frac{s}{P_{O-A}}}{(1+A_{o\_dc}\beta)\left[1+\frac{s}{(1+A_{o\_dc}\beta)P_{O-A}}\right]} \quad (5)$$

$$= \frac{1 + \frac{s}{BW_A}}{(A_{o\_dc}\beta)(1+\frac{s}{GBW})}$$

From the above equation, it can be seen that the loop gain $-3\ dB$ bandwidth is the zero point $Z_1$ of the PSRR transfer function, and the $GBW$ is the pole $P_1$ of the PSRR transfer function. When the power supply ripple frequency w exceeds $BW_{AA}$, the PSRR performance deteriorates. $BW_{AA}$ appears as a turning zero in the PSRR frequency response plot. Within the mid-frequency range, as $Z_{o\_reg}$ continues to increase, PSRR performance continues to deteriorate.

When w increases to $GBW$, since this point is the critical point where the feedback system fails, the closed-loop output impedance effect no longer exists, and the feedback system can no longer reduce the output impedance. The deterioration of the PSRR frequency response stops at the $GBW$ point, where (3) can be rewritten as:

$$PSRR_{f=GBW} = \frac{Z_o}{Z_o + r_{ds}} = \frac{R_1 + R_2}{R_1 + R_2 + r_{ds}} \approx 1 \quad (6)$$

As shown in (6), when the ripple frequency is near the unity gain bandwidth, the system's PSRR reaches its worst-case scenario, with the LDO responding to nearly all of the ripple

When the ripple frequency is outside the $GBW$, it is considered high-frequency. The equivalent capacitive reactance of the load $C_{out}$, $1/(C_{out} \times j\omega)$, significantly decreases, causing $Z_o$ to be primarily determined by capacitive reactance. The power supply ripple begins to weaken in the output response, and the PSRR shows an improving trend. Without neglecting the equivalent resistance of the load capacitor $R_{ESR}$, (3) can be modified as follows:

$$PSR_{f>GBW} = \frac{Z_o}{r_{ds} + Z_o} = \frac{1/sC_{out} + R_{ESR}}{r_{ds} + 1/sC_{out} + R_{ESR}}$$

$$= \frac{1 + \dfrac{s}{1/C_{out}R_{ESR}}}{1 + \dfrac{s}{1/C_{out}(R_{ESR} + r_{ds})}} \qquad (7)$$

As shown in (7), the PSRR has a pole $P_2 = 1/C_{out}(R_{ESR} + r_{ds})$ and a zero point $Z_2 = 1/C_{out}R_{ESR}$ in the frequency band other than GBW, and both frequency points are shown in Fig. 3, which verifies the accuracy of this equation.

Combining (4),(5),(6),(7), the low-frequency PSRR of the LDO, the PSRR performance inflection point, and the subsequent three frequency points are determined by the system's low-frequency loop gain, the loop gain $-3\ dB$, the loop unity gain frequency point $GBW$, the system's high-frequency output pole, and the high-frequency $ESR$ zero point, respectively.

In summary, the three-band frequency analysis method intuitively and clearly analyzes the PSRR performance influencing factors in a wide frequency domain, but the method is limited in that it does not take into account the influence of the power supply ripple conduction characteristics of the error amplifier and the bandgap reference itself; while the two-stage op-amp method does not analyze the PSRR in a wide frequency domain, but it gives the contribution of the various levels of PSRR to the overall PSRR, and the combination of the two methods can be used as a guide for the design of a high power supply ripple suppression LDO. The combination of the two methods can be used as a guide for the design of high supply ripple rejection LDOs

*D. LDO transient response research method*

When the load current or supply voltage changes abruptly, the LDO's output voltage may experience overshoot and undershoot due to its inherent closed-loop bandwidth and slew rate. Subsequently, the LDO uses its linear negative feedback system to restore the output voltage to a stable value. These processes are referred to as load transient response and linear transient response, respectively.

The transient response of an LDO is determined by the large-signal and small-signal characteristics of its loop. Overvoltage and recovery time are key metrics for optimizing transient performance, with the former determining the maximum change in the LDO's output voltage and the latter determining the time required for the output voltage to return to a stable value. This report will quantitatively analyze these two metrics based on the large-signal and small-signal response processes of the LDO loop, and discuss how to optimize transient response performance by coordinating the relationship between the error amplifier's bandwidth, slew rate, and power within the LDO loop. The relevant theories discussed here provide guidance for analyzing the specific simulation results of actual circuits.

As shown in Fig. 4, when the LDO driver transitions abruptly from light load $I_{L,min}$ to heavy load $I_{L,max}$ the output load capacitance $C_L$ responds first within the undershoot time $\Delta t_1$. The discharge of $C_L$ pulls the output voltage down, causing the error amplifier to enter a transition state and the feedback loop to fail. The large gate parasitic capacitance $C_{par}$ of the power transistor is limited by its slew rate and requires corresponding discharge time. As $C_{par}$ discharges, the gate voltage of the power transistor is pulled down, enhancing its drive capability to meet the drive requirements, and the slew stops at time $t_1$. Within $\Delta t_1$, the discharge time of $C_{par}$ dominates, so the downward slew process is a large-signal response.

$$\Delta V_{TR+} = \frac{I_{L,\max}}{C_L}\Delta t_1 + \Delta V_{esr} = \frac{I_{L,\max}}{C_L}\Delta t_1 + I_{L,\max}R_{ESR} \quad (8)$$

$$\Delta t_1 \approx \frac{1}{BW_{OUT}} + C_{par}\frac{\Delta V_G}{I_{S-}} \qquad (9)$$

The undershoot voltage $\Delta V_{TR+}$ and undershoot time $\Delta t_1$ are shown in (8) and (9), where $R_{ESR}$ is the equivalent internal resistance of $C_L$, $\Delta V_G$ and $I_{S-}$ are the gate adjustment voltage and error amplifier output negative slew rate current of the power transistor, respectively, and $BW_{OUT}$ is the closed-loop output pole. Increasing the slew rate current can reduce the slew response time, and smaller $C_{par}$ and $C_L$ can also reduce the slew response time, but this comes at the cost of increasing the slew voltage value. During the design process, especially for on-chip compensated LDOs, trade-offs must be considered.

After time $t_1$, the output current of the adjustment tube begins to charge the load capacitor $C_L$, causing the output voltage to rise. The error amplifier exits the transition state and begins normal operation, with the main loop feedback taking effect. The small-signal output feedback is subtracted from the reference voltage of the error amplifier, and the gate voltage of the power transistor is fine-tuned to adjust the output drive. After repeated adjustments through the feedback loop, the output voltage stabilizes after the recovery time $\Delta t_2$. The recovery process is a small-signal response, where $\Delta t_2$ is inversely proportional to the system's closed-loop $3dB$ bandwidth; the larger the bandwidth, the shorter the response recovery time. Due to the limited loop gain, the final stable value will be slightly less than the rated output voltage $\Delta V_{LDR}$.

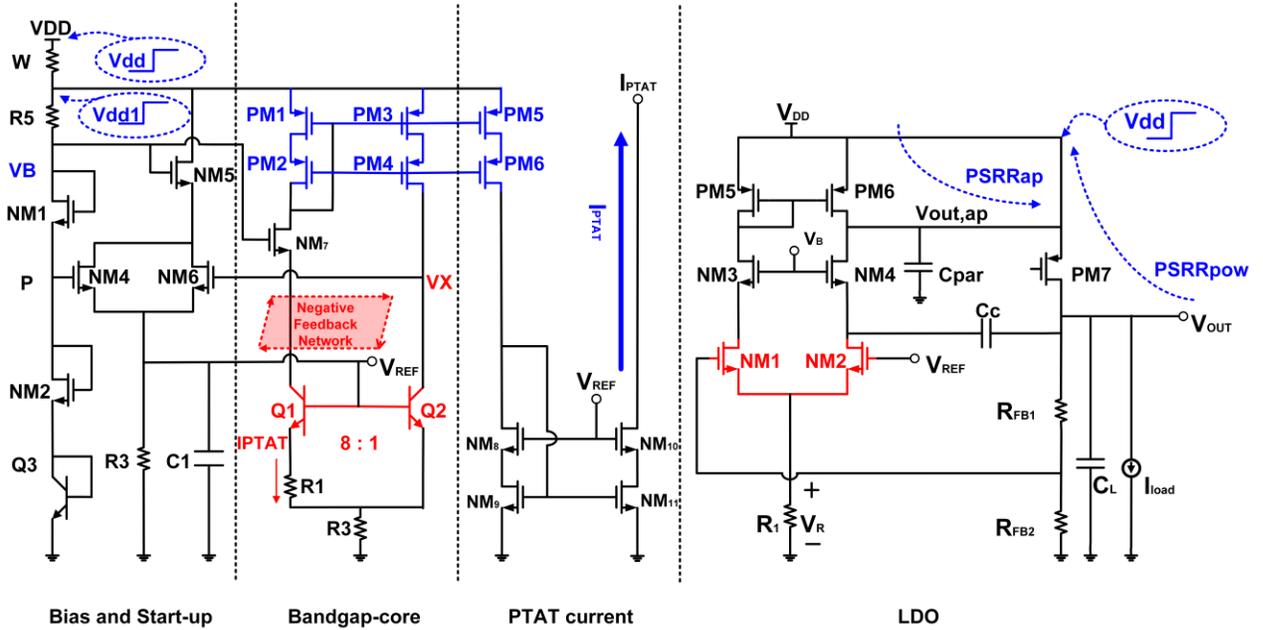

**Fig. 4**. Proposed LDO Overall Circuit.

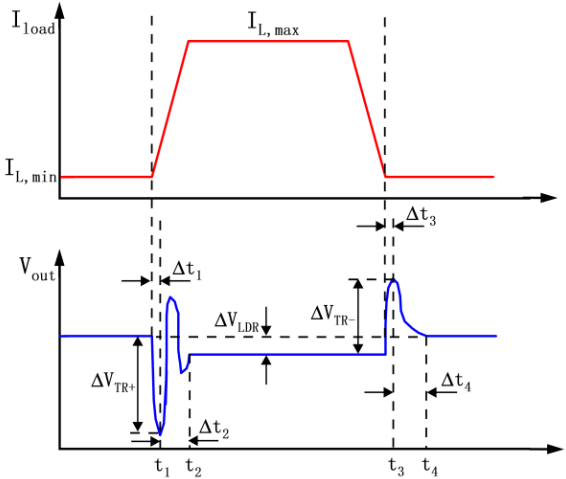

**Fig. 5**. Output Transient Response of LDO during Load Transient Changes.

When $I_{L,max}$ changes to $I_{L,min}$, the drive current first charges $C_L$, causing the output voltage to overshoot. The loop fails, and $C_{par}$ is limited by the slew rate, requiring a corresponding charging time. During the overshoot time $\Delta t_3$, $C_{par}$ charging still dominates, so the overshoot process is a large signal response.

$$\Delta V_{TR-} = \frac{I_{L,max}}{C_L}\Delta t_3 + \Delta V_{esr} \quad (10)$$

$$\Delta t_3 \approx \frac{I_{L,max}}{C_L}\frac{1}{BW_{OUT}} + C_{par}\frac{\Delta V_G}{I_{S+}} \quad (11)$$

The overshoot voltage $\Delta V_{TR-}$ and overshoot time $\Delta t_3$ are shown in (10) and (11), where $I_{S+}$ is the positive slew rate current of the error amplifier. The positive and negative slew rate currents are different, resulting in differences between the overshoot time and undershoot time.

When $C_{par}$ charges to the $t_3$ time, the power transistor enters the subthreshold region or the cutoff region edge, causing the rise to stop. The power transistor either does not provide or only provides a small amount of drive. Subsequently, $C_L$ begins to discharge from the feedback branch, lowering the output voltage, and the loop begins to respond. After the feedback loop is repeatedly adjusted through the recovery time $\Delta t_4$ to the moment $t_4$, the output is steady state and $\Delta t_4$ remains a small signal response.

$$\Delta t_4 \approx \frac{C_L}{I_X}\Delta V_{TR-} \quad (12)$$

The recovery time $\Delta t_4$ of the overshoot is shown in (12), where $I_X$ is the discharge current of the feedback branch. Above for the LDO load transient response process analysis, its linear transient response is also dynamic performance, the measure is still the response process of the output up and down stroke voltage value and the up and down stroke response time, the analysis process is similar to the load transient response, will not repeat.

IV. CIRCUIT AND SIMULATION RESULTS

The previous section provided a detailed analysis of the working principle of the LDO and the methods for studying its key performance parameters. This report will now present the specific LDO circuit implementation and validate the theoretical research through actual circuit and simulation results.

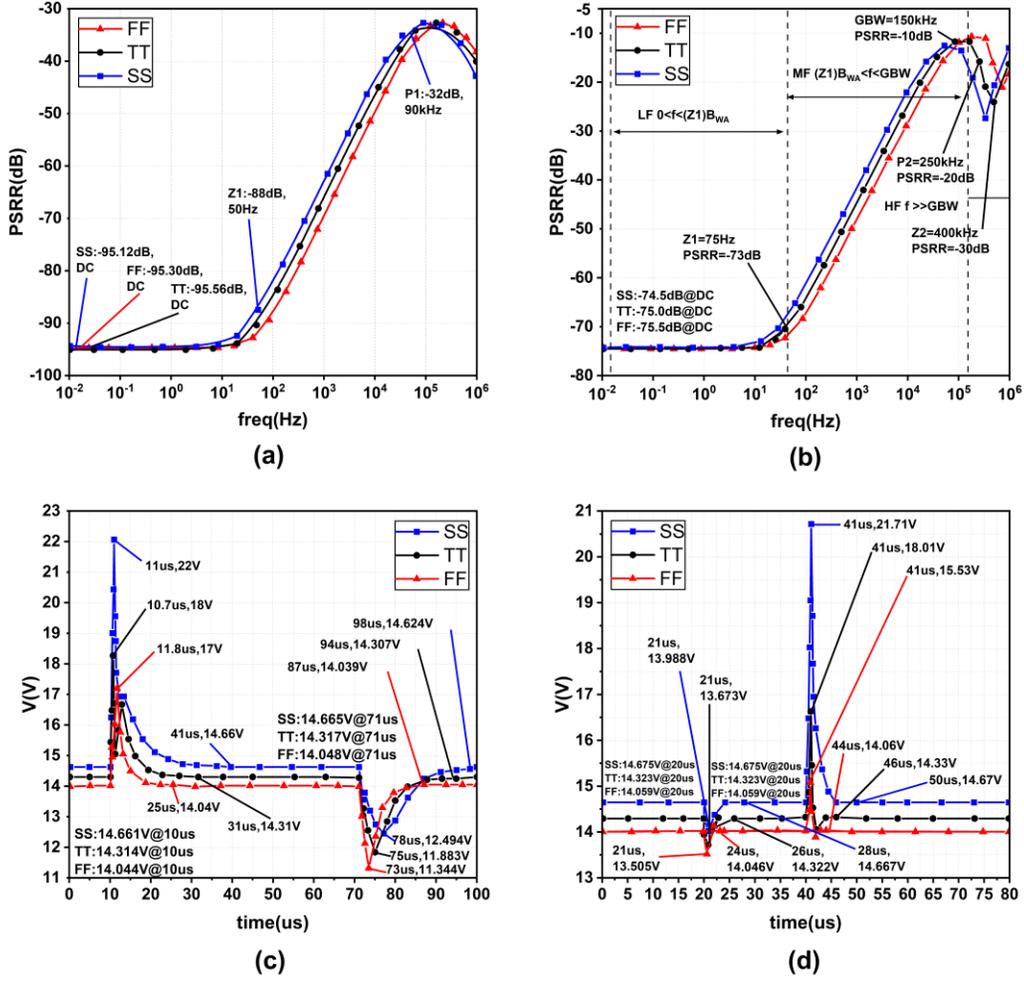

**Fig. 6**. Simulation Results: (**a**) Bandgap reference PSRR; (**b**) LDO PSRR; (**c**) LDO linear transient response; (**d**) LDO load transient response.

Based on the $CSMC\ 0.25\mu m\ BCD$ process, the LDO circuit designed using $Cadence\ Virtuoso\ EDA$ tools are shown in Fig. 4. The specific design approach of the LDO is not elaborated upon here. The report will primarily validate the practicality of the two-stage op-amp PSRR analysis method from a circuit-level perspective and verify the three-band PSRR analysis method and related theories on transient response overshoot voltage and recovery time based on the overall LDO circuit shown in Fig. 4 (This LDO design employs a mixed high-voltage and low-voltage transistor configuration, enabling normal operation within the $V_{DD}$ range of $4.5V$ to $60V$.)

To verify the effectiveness of the architecture, we validated the two analysis methods from the circuit layer and simulation perspectives, respectively.

*A. Two-stage op-amp PSRR analysis method verified at the circuit level*

In Fig. 4, if the gate and source of $PM_7$ have power supply noise of the same magnitude, the two noise paths can cancel each other out at the output $V_{out}$ (the common-source and common-gate gain amplitudes are the same, but their phases are opposite). Combining the cancellation concept,

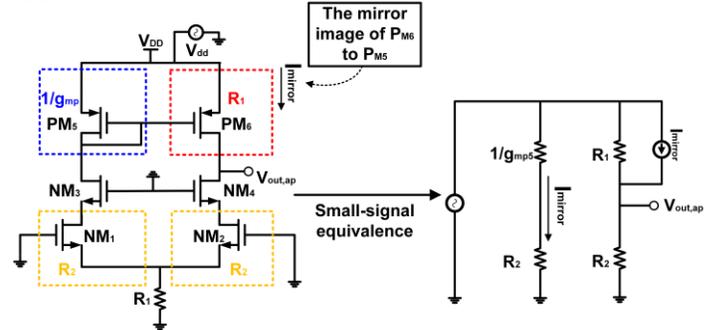

**Fig. 7.** The effect of power supply noise on the output of error amplifiers

Fig. 7 shows the effect of power supply noise $V_{dd}$ on the error amplifier output $V_{out,ap}$. Using the superposition theorem, we can list:

$$V_{out,ap} = V_{dd}(\frac{R_2}{R_1+R_2}) + I_{mirror}(R_1 \| R_2)$$
$$\approx V_{dd}(\frac{R_2}{R_1+R_2}) + \frac{V_{dd}}{R_2}(\frac{R_1 R_2}{R_1+R_2}) = V_{dd} \qquad (13)$$

As shown in (13), the error amplifier's output nearly losslessly transmits the power supply ripple, resulting in $PSRR_{EA} = 1$ in (2), which satisfies $-A_2 + PSRR_{POW} = 0$. Therefore, by adopting this main loop architecture, the effects of the last two terms in (3) that influence PSRR performance are canceled out [6]. In summary, the two-stage operational amplifier PSRR analysis method can be combined with actual circuit design to optimize the PSRR performance of an LDO, making it a practical method for verifying the feasibility of the loop architecture of the designed LDO circuit.

*B. Verify the three-band PSRR analysis method and transient response characteristics at the simulation result level*

*a. Bandgap Reference Circuit PSRR Simulation*

Under the conditions of $TT, FF, SS$ process Corner, $T = 25 °C$, and $V_{CC} = 48\ V$, the bandgap-based PSRR simulation is shown in Fig. 6(a) and we can see that under $SS, FF, TT$ process corner, the bandgap reference low-frequency PSRR is $-95.12\ dB$, $-98.29\ dB$, and $-95.56\ dB$, respectively.

*b. LDO Circuit PSRR Simulation*

Under the conditions of $TT, FF, SS$ process corner, $T = 25°C$, and $V_{CC} = 48\ V$, the LDO PSRR simulation results are shown in Fig. 6(b). As shown in the figure, the PSRR curves for each process corner conform to the trends predicted by the three-band PSRR theoretical analysis. The PSRR transfer curves shown in Fig. 3 and Fig. 6(b) are highly consistent. Taking the $TT$ process corner as an example: the low-frequency PSRR of the $14\ V$ LDO, which is of primary concern, is $-75\ dB$, indicating good power supply ripple suppression performance. The other frequency bands of interest in the design are as follows: -73dB@$Z_1$ (75Hz), -10dB@GBW (150kHz), -20dB@$P_2$ (250kHz), -30dB@$Z_2$ (400kHz).

It is worth noting that, under low-frequency conditions, as shown in Fig. 6(a) and Fig. 6(b), the overall PSRR value of the LDO is lower than that of the bandgap reference module.(2) can predict that, although the theoretical analysis of the LDO architecture designed in this report suggests that the last two terms of (2) should cancel each other out, some non-ideal factors in the actual circuit cause the difference between the last two terms to be negative. This negative value, when added to the PSRR value indicated by the bandgap reference, further results in the overall circuit's PSRR simulation result being slightly lower than that of the reference module.

*c. LDO Circuit Transient Response Performance Simulation*

Under $TT, SS, FF$ process conditions, $T = 25°C$, and heavy load 4 $mA$ conditions, the input voltage $V_{CC}$ changes from 16 $V$ to 26$V$ (16$V/\mu s$), stabilizes for a period of time, and then jumps back to 16$V$ (10$V/\mu s$). The linear transient response of the LDO is shown in Fig. 6(c).

Using the $TT$ process corner in Fig. 6(c) as an example, analyze the linear transient response of the LDO: Within 0~10$\mu s$, the input voltage $V_{in} = 16\ V$, and the output voltage is 14.314 $V$; At 11 $\mu s$, $V_{in}$ jumps to 26$V$(10$V/\mu s$), with an overshoot voltage of 4 $V$. After 20 $\mu s$, the output stabilizes, limited by the finite loop gain, with the output voltage slightly higher than the initial voltage at 14.317 $V$; At 71 $\mu s$, $V_{in}$ jumps to 16$V$(10$V/\mu s$), with a overshoot voltage of 2.434 $V$. After 19 $\mu s$, the output stabilizes, still limited by the finite loop gain, with the output voltage slightly lower than the initial voltage at 14.307 $V$.

Under the conditions of $TT$, $SS$, $FF$ process corner, $T = 25\ °C$, and $V_{CC} = 48\ V$, the simulation results of the LDO's load transient response are shown in Fig. 6(d).

Using the $TT$ process corner in Fig. 6(d) as an example, analyze the load transient response of the LDO: Within the 0~20$\mu s$ time interval, the LDO operates in a 50 $\mu A$ light load state, with an output voltage of 14.323 $V$; At 20 $\mu s$, the load current transitions from a 50 $\mu A$ light load state to a 4 $mA$ heavy load state in 1 $\mu s$, with a dropout voltage of 0.65 $V$. After 16 $\mu s$, the output voltage stabilizes, limited by the finite loop gain, and is slightly lower than the initial voltage at 14.322 $V$; At 40 $\mu s$, the load current jumps from 4 $mA$ heavy load state to 50 $\mu A$ light load state in 1 $\mu s$, with an overshoot voltage of 4 $V$. After 6 $\mu s$, the output stabilizes, still limited by the finite loop gain, with the output voltage slightly higher than the initial voltage at 14.33 $V$.

In summary, the linear/load transient response of the designed LDO aligns with the trend predicted by the simulated curve in Fig. 4, and the transient characteristics of the designed LDO are excellent (moderate overshoot and undershoot voltages, and fast stabilization time). This further validates the feasibility and stability of the proposed design method under actual process conditions.

Summary of the Report

This practical report first adopts a theoretical approach to systematically analyze the key factors influencing the PSRR and transient response performance of an LDO. Using a two-stage operational amplifier analysis method, a three-band PSRR analysis method, and transient response simulation processes, the PSRR and transient response performance of the LDO are analyzed. Subsequently, the selected process-designed LDO circuit and its simulation results were used to validate the relevant theories involved. The validation results at both the circuit level and simulation result level indicate that the research methods and techniques proposed in this report provide a feasible path for the integrated implementation of high-performance, wide-input-voltage LDOs, demonstrating practical applicability.

> REPLACE THIS LINE WITH YOUR MANUSCRIPT ID NUMBER (DOUBLE-CLICK HERE TO EDIT) <   8